\documentclass[reprint,amsmath,amssymb]{revtex4-1}
\usepackage{graphicx}
\usepackage{dcolumn}
\usepackage{bm}
\usepackage{color}

\begin{document}
\title{Acoustic Wavefront Manipulation: Impedance Inhomogeneity and Extraordinary Reflection}
\author{Jiajun Zhao$^{1,2}$, Cheng-Wei Qiu$^{1}$, Zhining Chen$^{1}$, and Baowen Li$^{2,3}$}
\affiliation{$^{1}$Department of Electrical and Computer Engineering, National University of Singapore, Singapore 117576, Republic of Singapore} \affiliation{$^{2}$Department of Physics and Centre for Computational Science and Engineering, National University of Singapore, Singapore 117546, Republic of Singapore}
\affiliation{$^{3}$NUS-Tongji Center for Phononics and Thermal Energy Science, School of Physical Science and Engineering, Tongji University, Shanghai 200092, People's Republic of China}
\date{\today}
\begin{abstract}
Optical wavefront can be manipulated by interfering elementary beams
with phase inhomogeneity. Therefore a surface allowing huge, abrupt
and position-variant phase change would enable all possibilities of
wavefront engineering. However, one may not have the luxury of
efficient abrupt-phase-changing materials in acoustics. This
motivates us to establish a counterpart mechanism for acoustics, in
order to empower the wide spectrum of novel acoustic applications.
Remarkably, the proposed impedance-governed generalized Snell's law
(IGSL) of reflection is distinguished from that in optics. Via the
manipulation of inhomogeneous acoustic impedance, extraordinary
reflection can be tailored for unprecedented wavefront manipulation
while ordinary reflection can be surprisingly switched on or off.
Our results may power the acoustic-wave manipulation and
engineering. We demonstrate novel acoustic applications by planar
surfaces designed with IGSL.
\end{abstract}
\maketitle
Refraction, a physical phenomena in classic optics, was recently
re-visited from the viewpoints of complex refractive index of a
bulky medium\cite{xx}, abrupt phase change of an interface
\cite{Capasso}, and diffraction theory for gratings
\cite{Smith-arXiv}. These works also shed light on the relation
between the reflection and incidence, interpreted as the generalized
Snell's law of reflection (GSL) \cite{Capasso}, which plays a novel
role on optical wavefront engineering and has resulted in promising
accomplishments \cite{Shalaev,Kang,Francesco,Huang,Genevet}. The
phase inhomogeneity, functioning as the underlying principle of
optical GSL, metamorphoses the original (ordinary) reflection into
the anomalous reflection through abrupt phase changes obtained at
different positions of the interface made of thin metallic
nanoantenna array. Fundamental physics may be explained by phase
antenna array \cite{Engheta,Balanis,ChenZN}.

GSL is based on Fermat's principle and hence the law should hold for
various wave types: electromagnetic and acoustic waves. But one
cannot analogously translate the GSL from electromagnetics to
acoustics, since the naturally available materials which can give
the abrupt phase change to the wave are limited in acoustics. Hence,
the luxury of using metallic meta-surface \cite{Capasso,Shalaev} to
fulfill the phase control is no more available, and it is necessary
to establish a distinct principle to manipulate the acoustic waves.
Our findings also reveal acoustic GSL is more complicated than that
for electromagnetics. In addition, the polarization of the anomalous
reflection controlled by electromagnetic GSL discords with that of
the incidence, though this is not explicitly presented in the law of
electromagnetic GSL \cite{Capasso}.

In this Letter, we establish the framework of acoustic wavefront
manipulation by resorting to specific acoustic impedance (SAI)
\cite{Blackstock} inhomogeneity, rather than the phase control as
Refs. \cite{xx,Capasso} in optical regime. More specifically, the
inhomogeneous SAI will generally give rise to one ordinary
reflection (analogous to anomalous reflection in \cite{Capasso}
controlled by GSL) and one extraordinary reflection (uniquely
pronounced in impedance-governed generalized Snell's law of
reflection in acoustics, i.e., IGSL).

In optical GSL \cite{Capasso}, the so-called anomalous reflection
actually corresponds to the situation when the original ordinary
reflection is deflected toward a ``wrong'' direction governed by
GSL. On the contrary, the ordinary reflection in IGSL cannot be
altered by an acoustic SAI interface, but IGSL can provide insight
to the design of SAI interface so as to ``turn off'' the ordinary
reflection. Moreover, the extraordinary reflection governed by IGSL
is an additionally unique component in acoustic cases, which can be
``geared'' along arbitrary directions in principle with vanishing
ordinary reflection simultaneously. Therefore our proposed IGSL can
lead to richer effects and applications in acoustics.

\begin{figure}
    \centering\includegraphics[scale=0.25]{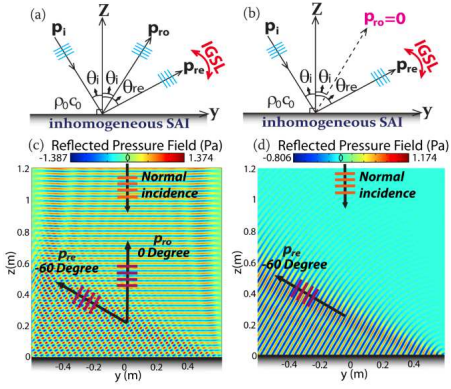}
\vspace{-2ex} \caption{\label{f1} Schematic of IGSL. $p_i$,
$p_{ro}$, and $p_{re}$ denote the incident acoustic wave, ordinary
reflection, and the extraordinary reflection. $\theta_i$ is the
incident angle of $p_i$. (a) For a flat interface with an
inhomogeneous SAI, the angle of $p_{ro}$, i.e., $\theta_{ro}$, is
not influenced, while $p_{re}$ occurs simultaneously and the angle
$\theta_{re}$ is controlled by IGSL. (b) If SAI is properly
controlled, $p_{ro}$ can be nulled. Ultrasound with unit amplitude
and $\omega=300Krad/s$ is normally impinged upon SAI surfaces from
water. (c) The SAI along the flat surface $z=0$ leads to both
$p_{ro}$ and $p_{re}$ when an arbitrary $A$ is chosen in
Eq.(\ref{Azn}). (d) A particular SAI inhomogeneity is chosen
according to Eq.(\ref{zn}). $\psi (y)=-100\sqrt 3y$ is selected
throughout.}
\end{figure}

The impedance ${Z_n}$ of the flat SAI surface is expressed in a complex form $A{e^{ - i\psi (y)}}$, where $A$ is the amplitude and $\psi (y)$ is the phase angle along the surface. Since the SAI is inhomogeneous, both the real and the imaginary parts of ${Z_n}$ change spatially. The intensity of an acoustic wave depends on the real part of a SAI and the root-mean-square acoustic velocity ${{\bf{v}}_{rms}}$, i.e., $I = {\mathop{\rm Re}\nolimits}({Z_n})v_{rms}^2$. Therefore, it is reasonable to set the real part of the SAI as a constant so that only one variable determines the intensity. In this connection, one may consider ${Z_n} = A{e^{ -i\psi (y)}}/\cos \psi (y)$, since $A$ is an arbitrary value. Now the real constant $A$ in the new form of ${Z_n}$ corresponds to the constant resistance of the SAI. After some algebraic treatment in the phase angle for the convenience of derivation, we thus obtain \begin{equation}\label{Azn}
{Z_n}(y,\omega ) = A\frac{1}{{\cos [\psi (y)/2]}}{e^{ - i\psi (y)/2}},
\end{equation}
in which case both the ordinary and extraordinary reflections exist for a general $A$. Note that $\omega$-dependency on the right hand side of Eq.(\ref{Azn}) is included into $\psi (y)$. According to our derivation provided in Supplement Materials (Section I), the directivity factor governing the extraordinary reflection is
\begin{equation}\label{dir}
\Psi ({\theta _{re}},\omega ) = \int_{ - \infty }^\infty{\frac{{{\rho _0}{c_0}}}{{2A}}} {e^{i\psi (y)}}{e^{i\frac{\omega}{{{c_0}}}(\sin {\theta _i} - \sin {\theta _{re}})y}}dy,
\end{equation}
where $\rho _0$ and $c_0$ represent the density of the ambient medium and the speed of sound in the upper space in Fig.~\ref{f1}. The integration of Eq.~(\ref{dir}) consequently results in a Dirac delta function
\begin{equation}\label{delta}
\Psi ({\theta _{re}},\omega ) = \frac{\rho_0c_0}{2A}\delta [{k_0}y(\sin {\theta _i} -\sin {\theta_{re}} ) +\psi (y)],
\end{equation}
which makes sense only for ${k_0}y(\sin {\theta_{re}} - \sin{\theta_i} ) = \psi (y)$. $k_0=\omega/c_0$ stands for the wave number. Hence the relation between the incident angle and the angle of extraordinary reflection satisfies:
\begin{equation}\label{IGL}
k_0[\sin {\theta_{re}}  - \sin {\theta _i}] =d\psi(y)/dy.
\end{equation}
Although IGSL's form seems similar to GSL, its physical origin and the meaning of $\psi(y)$ are dramatically different from those of optical GSL \cite{xx,Capasso,Shalaev}. Moreover, IGSL only serves to redirect the extraordinary reflected wave arbitrarily, with no influence on the direction of ordinary reflection. In the mean time, Eq.~(\ref{IGL}) sheds a light to an extreme angle (similar to critical angle):
\begin{equation}\label{ca}
{\theta _e} = \left\{ \begin{array}{l}
\arcsin ( - 1 - \frac{1}{{{k_0}}}\frac{{d\psi (y)}}{{dy}}){\rm{,if}}\frac{{d\psi (y)}}{{dy}} < 0\\
\arcsin ( + 1 - \frac{1}{{{k_0}}}\frac{{d\psi (y)}}{{dy}}){\rm{,if}}\frac{{d\psi (y)}}{{dy}} > 0
\end{array} \right.,
\end{equation}
above which extraordinary reflection becomes evanescent.
Equation~(\ref{ca}) holds only if $- 1 \le 1-\frac{1}{{{k_0}}}\left|
{\frac{{d\psi (y)}}{{dy}}} \right| \le 1$. Otherwise, extraordinary
reflection does not propagate in the upper space.

Usually, both $p_{ro}$ and $p_{re}$ will coexist as shown in
Fig.~\ref{f1}(a), suggesting \emph{double} reflection, while IGSL
only controls the angle of extraordinary reflection ${\theta
_{re}}$. Hence, it is interesting to eliminate $p_{ro}$ in
Fig.~\ref{f1}(b), by means of a particularly selected value of $A$
in Eq.~(\ref{Azn}). Our elaboration in Supplement Materials (Section
I) suggests that $A=({\rho _0}{c_0})/(2\cos {\theta _i})$ can make
$p_{ro}$ vanished, i.e., the ordinary reflection is switched off, as
shown in Fig.~\ref{f1}(b,d). Corresponding SAI of the flat surface
is
\begin{equation}\label{zn}
{Z_n}(y,\omega ) = \frac{{{\rho _0}{c_0}}}{{2\cos {\theta_i}}}\frac{1}{{\cos [\psi (y)/2]}}{e^{ - i\psi (y)/2}}.
\end{equation}
Only at such condition, acoustic IGSL behaves as an exact
counterpart of optical GSL whereas the mechanisms are completely
different.

Supposing the gradient of $\psi(y)$ along the flat interface in
Eq.~(\ref{zn}) is constant, it can be predicted from Eq.(\ref{IGL})
that the wavefront of extraordinary reflection will propagate as a
plane acoustic wave, independent of the location $y$. We selected
water (${\rho _0} = 1kg/{m^3}$; ${c_0} = 1500m/s$ \cite{Blackstock})
in the upper space, $\omega=300Krad/s$ as the circular frequency,
$e^{-i{k_0}z}$ as the normal incident plane ultrasound, and a linear
form of $\psi (y)=-100\sqrt 3y$ in Eq.~(\ref{zn}).

The angle of extraordinary reflection is theoretically found to be
${-60^ \circ}$ by IGSL, validated by our simulation in
Fig.~\ref{f1}(d). The ordinary one is thoroughly suppressed thanks
to the specific $A$ chosen according to Eq.(\ref{Azn}). The vanished
ordinary reflection is achieved by adjusting the SAI so as to make
the incident angle meets the angle of intromission
\cite{Blackstock}. In contrast, in Fig.~\ref{f1}(c), the same
parameters are kept except for another option for $A$, whose value
is arbitrarily taken to be $2{\rho_0}{c_0}$. It clearly shows that
the ordinary reflection occurs, and meanwhile the extraordinary one
still keeps the same angle ${-60^\circ}$, verifying the stability of
our theoretical formulation.

\begin{figure}
    \centering\includegraphics[scale=0.35]{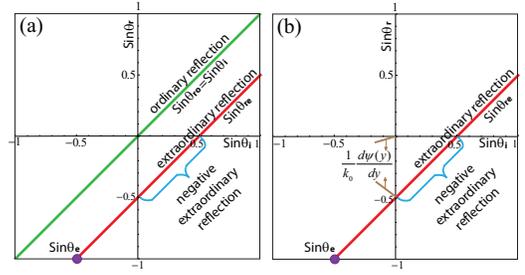}
\vspace{-2ex} \caption{\label{f2} Reflection angles $\sin
{\theta_{ro,re}}$ versus $\sin {\theta _i}$ when $k_0=10rad/m$ and
$\psi (y)=-5y$. Ordinary (green line) and extraordinary reflections
(red line) emerge simultaneously in (a). In (b), only extraordinary
component is present for the same parameters of (a) except $A$. The
purple dot denotes $\sin {\theta_e}$ in Eq.~(5).}
\end{figure}

Figure~\ref{f1}(d) has suggested the possibility of negative
extraordinary reflection, which is verified for oblique incidence in
Fig.~\ref{f2}. In Fig.~\ref{f2}(a), because of the inhomogeneous SAI
and the arbitrary $A$ in Eq.~(\ref{Azn}), both ordinary reflection
and extraordinary reflections occur. Fig.~\ref{f2}(b) depicts the
same situation except for the ordinary reflection being switched-off
as a result of the specifically chosen $A$ according to
Eq.~(\ref{zn}), while the red line stays put as that in
Fig.~\ref{f2}(a). The blue braces along the red line (extraordinary
reflection) represent the region of negative reflection. It is
noteworthy that extraordinary reflection does not exist if an
incident angle is beyond the extreme angle $\theta_e=-30^\circ$ as
described in Eq.~(\ref{ca}), corresponding to the purple dots in
Fig.~\ref{f2}.

\begin{figure}
    \centering\includegraphics[scale=0.30]{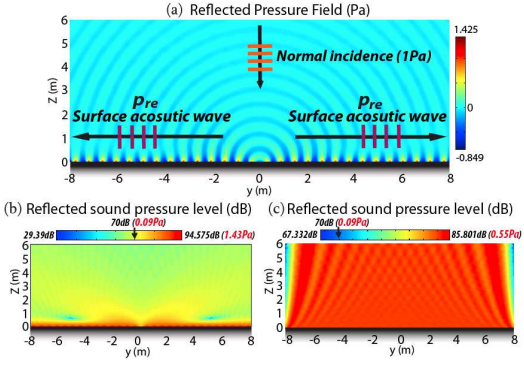}
\vspace{-2ex} \caption{\label{f3} Conversion of PAWs to SAWs via SAI
interface. The PAW with unit amplitude and $\omega=15Krad/s$ is
normally incident from water. Only reflected acoustic pressure is
plotted. (a) The SAI of Eq.(\ref{zn}) is set to be $\psi (y)=-11y$
for $y<0$ and $\psi(y)=11y$ for $y>0$. SAWs are bifurcated at the
origin and confined near the surface when propagating. (b) The
reflected sound pressure level of (a). (c) The reflected sound
pressure level when a homogeneous SAI as in Eq.(\ref{zn}) with
$\psi(y)=11$ is adopted instead.}
\end{figure}

To demonstrate IGSL's capability of designing novel acoustic
devices, we proposed a SAI surface which can convert a propagating
acoustic wave (PAW) to a surface acoustic wave (SAW) in
Fig.~\ref{f3}. It can be verified from IGSL that the extreme angle
$0^ \circ$ in Eq.(\ref{ca}) demands $\psi (y)=\pm 10y$. Therefore,
one can set the SAI of Eq.~(\ref{zn}) slightly beyond that extreme,
e.g., $\psi(y)=-11y$ for $y<0$ and $\psi (y)=11y$ for $y>0$ is set
along the flat interface and symmetric with respect to the $z$ axis.
In Fig.~\ref{f3}(a), the arrows with purple crossbars label the
paths of the bidirectional surface acoustic waves, which are caused
by the evanescent extraordinary reflection owing to the
inhomogeneous SAI interface, with a bit diffraction implicating the
ideally perfect conversion. In contrast, if one uses an
inappropriate SAI as in Eq.(\ref{zn}) with $\psi (y)=11$ along the
flat surface (the homogenous SAI does not generate extraordinary
reflection; only ordinary reflection occurs), the reflected sound
pressure level in Fig.~\ref{f3}(c) is almost uniformly spread over
the space.

Figure~\ref{f3}(b) clearly demonstrates that the acoustic field is
well confined in the region close to the interface and attenuated
quickly to around $0 Pa$ away from the interface, revealing the
nearly perfect conversion. Interestingly, it shows in \cite{Zhou}
that the meta-surface (H-shaped design) provides an extra momentum
to overcome the momentum mismatch from a propagating electromagnetic
wave to a surface one, in which a mushroom structure in addition to
the fractal surface played a dispensable role to continuously couple
out the surface electromagnetic wave generated by the fractal. This
prevents the propagating electromagnetic waves from being reflected
back to the upper space. Hence, our PAW-SAW conversion in acoustics,
originating form a distinguished mechanism, is differentiated from
\cite{Zhou} in terms of physical principle and application domain.

From Fig.~\ref{f3}, one may notice that such technology can function
as an alternative invisibility acoustic cloak by trapping the
acoustic field in the vicinity of the coating, resulting in much
lower signal strength of the reflection. It may pave the avenue to
the large size acoustic invisibility since it is only dependent on
the surface technique instead of wave-interaction based metamaterial
acoustic cloaking \cite{Shu}.

\begin{figure}
    \centering\includegraphics[scale=0.22]{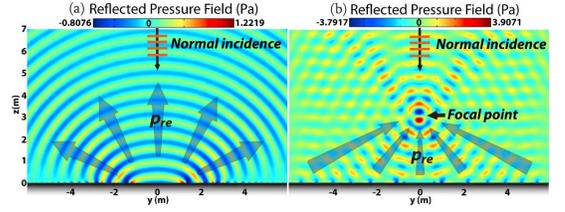}
\vspace{-2ex} \caption{\label{f4} Wavefront metamorphosis via SAI
interface. A plane acoustic wave of $\omega=15Krad/s$ is normally
incident from water. Only reflected acoustic pressure is plotted.
(a) The SAI of Eq.(\ref{zn}) with $\psi (y)=2.5y^2$ is set along the
flat surface. Extraordinary reflection, represented with the
translucent arrows, diverges in a different wavefront. (b) The SAI
of Eq.(\ref{zn}) with $\psi (y)=-10\left( {\sqrt {{y^2} +
{3^2}}-3}\right)$ is set. Extraordinary reflection converges to a
focal point in the two-dimensional case.}
\end{figure}

If one considers the PAW-SAW conversion to be a creation of acoustic
cognitive deception, our IGSL can also metamorphose acoustic
pressure fields everywhere through SAI manipulation. This deception
effect is obtained by manipulating plane wavefronts into wavefronts
generated by a virtual source or focusing illumination, governed by
the control of extraordinary reflection by means of IGSL. Under
these scenarios, we need to consider nonlinear form of $\psi (y)$,
which means the right hand side of Eq.(\ref{IGL}) relies on $y$. New
phenomena can thus be expected when the angle of extraordinary
reflection becomes spatially varying.

It is found that the acoustic deception can indeed be created by
IGSL, e.g., a quadratic SAI $\psi(y)=2.5y^2$ in Eq.~(\ref{zn})
resulting in ordinary reflection eliminated. Correspondingly, the
angle of extraordinary reflection in Fig.~\ref{f4}(a) is a
position-dependent function $\sin \theta_{re} =0.5y$, functioning
only within the region of $-2\le y\le2$ of the SAI interface. Within
this particular region, the extraordinary reflected waves will fan
out into the upper space as demonstrated in Fig.~\ref{f4}(a), which
verifies our theoretical prediction. Beyond $-2\le y\le2$, no
propagating extraordinary reflection can be excited owing to
Eq.(\ref{ca}). Therefore, IGSL can be employed to camouflage a flat
surface as if there were an emitting source at the origin instead of
the physical planar interface. The dual effect by camouflaging
curvilinear interfaces by a virtually flat one, by manipulating the
convex wavefronts into planar wavefronts, was reported in plasmonic
regime \cite{Renger}.

Furthermore, the SAI can be manipulated to let acoustic waves
reflected by a planar interface be focused as well. In optics, a
flat lens with metallic nanoantennas of varying sizes and shapes can
consequently converge the transmitted light to a focal point
\cite{Kang,Francesco}. It is worth noting that the optical focal
controlled by optical generalized Snell's law of refraction is on
the other side of incoming light, i.e., on two sides of the flat
surface \cite{Kang,Francesco} in the transmission mode. On the
contrary, in acoustics, we employed an inhomogeneous SAI flat
surface to focus the extraordinary reflection wave, in the
reflection mode suggested by IGSL developed in this Letter.

This \emph{ipsilateral} focusing, as demonstrated in
Fig.~\ref{f4}(b), is thus obtained in the planar geometry in
acoustics for the first time. In Eq.~(\ref{zn}), a hyperbolic form
was set $\psi (y) =- {k_0}\left( {\sqrt {{y^2} + {f^2}}-f}\right)$
($f$ being the given focal length \cite{Hecht}) for the SAI of the
flat interface. Extraordinary reflections from different angles
constructively interfere at the ipsilateral focal point, as if the
waves are emerging from a parabolic surface. The parameters in
Fig.~\ref{f4}(b) are the same as those in Fig.~\ref{f4}(a) except
for the specific hyperbolic SAI form $\psi (y)=-10\left(
{\sqrt{{y^2} + {3^2}} -3}\right)$, with the designed focal point at
$(y =0,z = 3)$ and the ordinary reflection suppressed. In addition,
the simulated acoustic pressure at the focal point is well confined
at $(y = 0,z = 3)$.

Interestingly, the imaging on the same side was only possibly
presented in \cite{Cui} for electromagnetic waves, which demands
strong chiral materials filled in the whole upper space and the
propagating wave excited inside the filling chiral material. The
same-side imaging is only a partial imaging, i.e., only one
circularly polarized wave being imaged and the other circularly
polarized wave being reflected ordinarily. In addition, it is
usually challenging to get high chirality. In acoustics, we do not
have the luxury of finding strong chiral materials and our
ipsilateral imaging is achieved by translating all the stringent
requirements of the half-space chiral materials into an
inhomogeneous impedance surface, and this phenomenon is independent
of circular polarization status.

Although we have shown robust capabilities of IGSL in obtaining
novel acoustic applications, it is still far from the theoretical
boundaries of what IGSL can be used to achieve. Nevertheless, an
equally important thing is to provide feasible schemes to realize
SAI by acoustic elements. In this connection, one representative
example is suggested for Eq.~(\ref{Azn}). Expanding Eq.~(\ref{Azn})
leads to $A - iA\tan [\psi (y)/2]$. In order to enhance the
reflection and reduce the absorbtion, it is imperative to have much
larger imaginary part than the real part, i.e., $\tan [\psi (y)/2]$
being so large that the real part of SAI can be ignored.

One method to realize the large $- iA\tan [\psi (y)/2]$ can be approached by
using hard-wall tubes with a pressure-release termination
\cite{Blackstock}, i.e., $- i{\rho _t}{c_t}\tan (\omega d/{c_t})$
where $\rho_t$ and $c_t$ are the density and the sound speed of the
medium in the tube respectively, and $d$ denotes the length of the tube. The
linear change of $d$ will result in the linear change of $\psi(y)/2$
correspondingly. If the range of $d$ leads to large $\tan[\psi
(y)/2]$ and the tubes with linearly increasing lengths are
juxtaposed perpendicular to one surface, particular SAI inhomogeneity
can be realized approximately by impedance discontinuity.

To conclude, the impedance-governed generalized Snell's law of reflection was
established for novel manipulations of acoustic wavefronts. It is
inspired by GSL which exploits abrupt-phase-changing materials in
optics, but the fundamental principle of IGSL completely differs
from that of GSL. IGSL can turn off ordinary reflection and control
extraordinary reflection in acoustics. GSL is only analogous to such
special situation of IGSL. We not only theoretically demonstrated
interesting manipulations of acoustic wave but also provided
insightful realization schemes for engineers. In addition, the cross
polarization is not present in acoustics. As a few examples, we
demonstrated acoustic PAW-SAW conversion, acoustic disguise, acoustic planar lens, and acoustic ipsilateral imaging. These novel
acoustic effects will inspire new technologies and devices on
acoustic wave engineering and manipulation, leading to unprecedented
applications.

We thank Prof.~Andrea Fratalocchi and Prof. Nanfang Yu for
stimulating discussion and advices.

\end{document}